# New scientific results with SpIOMM: a testbed for CFHT's imaging Fourier transform spectrometer SITELLE

L. Drissen[*a], A. Alarie[a], T. Martin[a], D. Lagrois[a], L. Rousseau-Nepton[a], A Bilodeau[a]
C. Robert[a], G. Joncas[a] and J. Iglesias-Páramo[b]

[a]Dépt. de physique, de génie physique et d'optique, Université Laval, Québec, Qc, Canada G1K 7P4
and Centre de recherche en astrophysique du Québec (CRAQ)
[b] Instituto de Astrofísica de Andalucía - CSIS

## ABSTRACT

We present new data obtained with SpIOMM, the imaging Fourier transform spectrometer attached to the 1.6-m telescope of the Observatoire du Mont-Mégantic in Québec. Recent technical and data reduction improvements have significantly increased SpIOMM's capabilities to observe fainter objects or weaker nebular lines, as well as continuum sources and absorption lines, and to increase its modulation efficiency in the near ultraviolet. To illustrate these improvements, we present data on the supernova remnant Cas A, planetary nebulae M27 and M97, the Wolf-Rayet ring nebula M1-67, spiral galaxies M63 and NGC 3344, as well as the interacting pair of galaxies Arp 84.

**Keywords:** Fourier transform spectroscopy, hyperspectral imagery, planetary nebulae, supernova remnants, galaxies

## 1. INTRODUCTION: SPIOMM, A PROTOTYPE FOR SITELLE

SpIOMM (**Sp**ectromètre **I**mageur de l'**O**bservatoire du **M**ont-**M**égantic) is an imaging Fourier transform spectrometer attached to the bonnette of the 1.6-m telescope of the Mont Mégantic Observatory (OMM) in southern Québec. It is capable of obtaining spectra in selected wavelength bands of the entire visible spectrum (from 350 to 850 nm) of every light source in a 12 arcminute field of view. The spectral resolution is variable, depending on the need of the observer, from R = 1 (broad-band image) to R = 25 000. In practice however, SpIOMM has been almost exclusively used at typical resolutions of R ~ 1000 - 2000 in conjunction with medium-band filters: SDSS u' (340 - 410 nm) and the tailor-made Spiomm-B (475 - 515 nm), Spiomm-V (545 - 635 nm) and Spiomm-R (645 - 680 nm). Spatial resolution is limited by the seeing (typically ~ 1.5 - 2"). The dual input, dual output design of SpIOMM ensures that virtually every photon collected by the telescope reaches the detector and is analyzed; a by-product of the spectral data cubes is therefore a very deep panchromatic image of the targets; however, only one output port was equipped with a detector until recently. Its early development phase and first science results were presented by Grandmont, Drissen and collaborators [1 - 5].

We have recently significantly improved SpIOMM in many ways that have allowed us to increase its capabilities:

- A fan-cooled Apogee Alta U3041 camera, with a 2k x 2k blue-sensitive CCD, was installed in the second output port of the interferometer. This not only doubles the signal detected by the instrument, but allows us to correct the interferograms for variations in sky transparency during the acquisition of a data cube. Although the CCD's quantum efficiency is higher in the blue range (350 - 450 nm) than that of the camera installed on the first output port (a 1340 x 1300 Princeton Instrument, LN2-cooled camera), its much higher dark current noise caused by its higher operation temperature (rarely below -45 C, and variable with an amplitude of a few degrees) is a handicap for faint targets. Also, vibrations caused by the cooling fans introduces noise in the metrology signal, which slightly decreases the modulation efficiency of the interferometer.
- Its structure and alignment were improved, making SpIOMM much less sensitive to external vibrations such as wind-shake. Whereas losses of the metrology signal were quite frequent in the past (resulting in lost data), they are almost non-existent today. This stabilizes the modulation efficiency at all wavelength, but particularly in the near-UV. SpIOMM is now a very stable instrument.

---

[*] ldrissen@phy.ulaval.ca; phone 1 418 656 2131 x-5641; fax 1 418 656 2040

- A totally new data reduction software, ORBS (see Martin et al., these proceedings), was developed by Thomas Martin. It takes into account the second camera, uses better algorithms for sky correction and is now fully automatized. The development of ORBS was, in part, motivated by the incoming commissionning of SITELLE at the CFHT (Grandmont et al., these proceedings).

All these improvements allowed us to study fainter objects, such as the external HII regions of spiral galaxies, observe faint emission lines in Galactic nebulae, detect absorption lines in stars and external galaxies and target emission lines in the near-ultraviolet. The data presented in the next section are shown to emphasize the capabilities of SpIOMM; a deeper analysis of their significance will be presented elsewhere.

## 2. RECENT SCIENCE RESULTS

**2.1 Targeting the third dimension: a 3D view of the Cassiopeia A supernova remnant**

Supernova remnants, which are the remains of the explosion of massive stars, display five strong emission lines in the 650 – 680 nm wavelength range of various intensities and Doppler shifts. While the stellar material in the oldest ones, such as NGC 6992 ( ~ 10 000 years old) , is now pretty well mixed with the interstellar medium and has considerably slowed down, the youngest supernova remnants of the Milky way, such as the Crab nebula (1054 AD) or Cassiopeia A (~ 1667 AD) are still expanding at very large velocities. Results for the Crab nebula are presented in Charlebois et al.[5]. Cas A has been observed with SpIOMM across the visible range, but we present here the most interesting results based on the red data cube. As shown in Figure 1, spectra of individual pixels along the nebula are very complex due to the overlapping of filaments moving of different chemical composition moving at different Doppler velocities: in some regions, up to three rapidly expanding filaments overlap with almost stationary material.

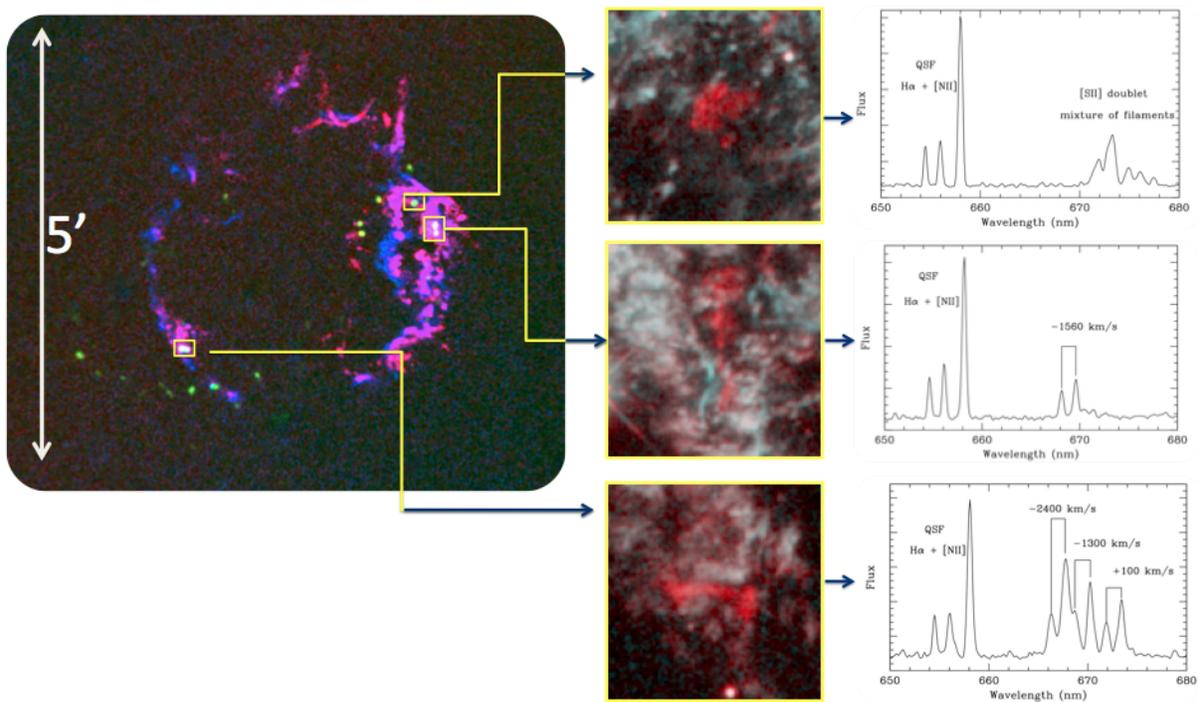

Fig. 1 (Left) - Doppler map of the supernova remnant Cas A. Blue and red colors are associated with Doppler-shifted [SII] 6717,6731 lines of the expanding filaments (amplitude of the Doppler shift ~ 8000 km/s). The green blobs are the nitrogen-rich Quasi-Stationary Floculi (QSF) emitted by the massive star prior to the supernova explosion. (Middle) - Hubble Space Telescope WFPC2 color-composite images (4" x 4") of three QSF superimposed on expanding filaments, obtained from a combination of F450W (blue) and F675W (red) images; the QSF stand out as red features. (Right) Spectra of the same QSF extracted from the data cube. Hα and [NII], almost at rest wavelength, from the QSF themselves are seen along with Doppler-shifted [SII] lines from multiple expanding filaments along the line of sight.

We have developed an algorithm that allows us to deconvolve the complex spectra and obtain a 3D view (the third dimension being the line-of-sight velocity), in all of the strongest emission lines (see [5]). In this 3D view, the nitrogen-rich Quasi-Stationary Floculi clearly stand out (left-hand panel of Figure 1).

We have also used *Hubble Space Telescope* archival images of Cas A obtained at different epochs during the last decade to determine the tangential velocity of the filaments, which, combined with the line-of-sight velocity from SpIOMM, provides a complete 3D velocity map of the nebula. Again, QSF stand out as stationary blobs of material, visible only in images obtained with the F675W filter (where the strong [NII] 654.8 nm is present), superimposed on rapidly-moving filaments.

## 2.2 Targeting faint lines: an hyperspectral view of the planetary nebula M27

Electron densities and temperatures have long been investigated in planetary nebulae and other HII regions. Reliable estimates are critical in order to obtain a clear picture of their chemical composition and overall abundances. Nonetheless, many problems and questions still exist. Abundance differences are revealed by comparing recombination to collisionally excited lines in gaseous nebulae. A few reasons for such discrepancies have been mentioned in the litterature (see, for instance, Mathis et al. 1998[6]). Among them, large fluctuations in electron densities and/or temperatures are commonly mentioned.

We have undertaken a survey of extended planetary nebulae with SpIOMM in order to: (1) investigate the impact of shock excitation in the excited zone, (2) extract bidimensional maps of the electron densities and temperatures using all line-ratio diagnostics provided by our data set, and (3) discuss their apparent morphology as from different ionic transitions (i.e., wavelengths).

M27 was an obvious early target because its extent almost completely fills SpIOMM's field of view and because it has been extensively studied with standard, long-slit spectrographs. We have therefore obtained data cubes with three filters.

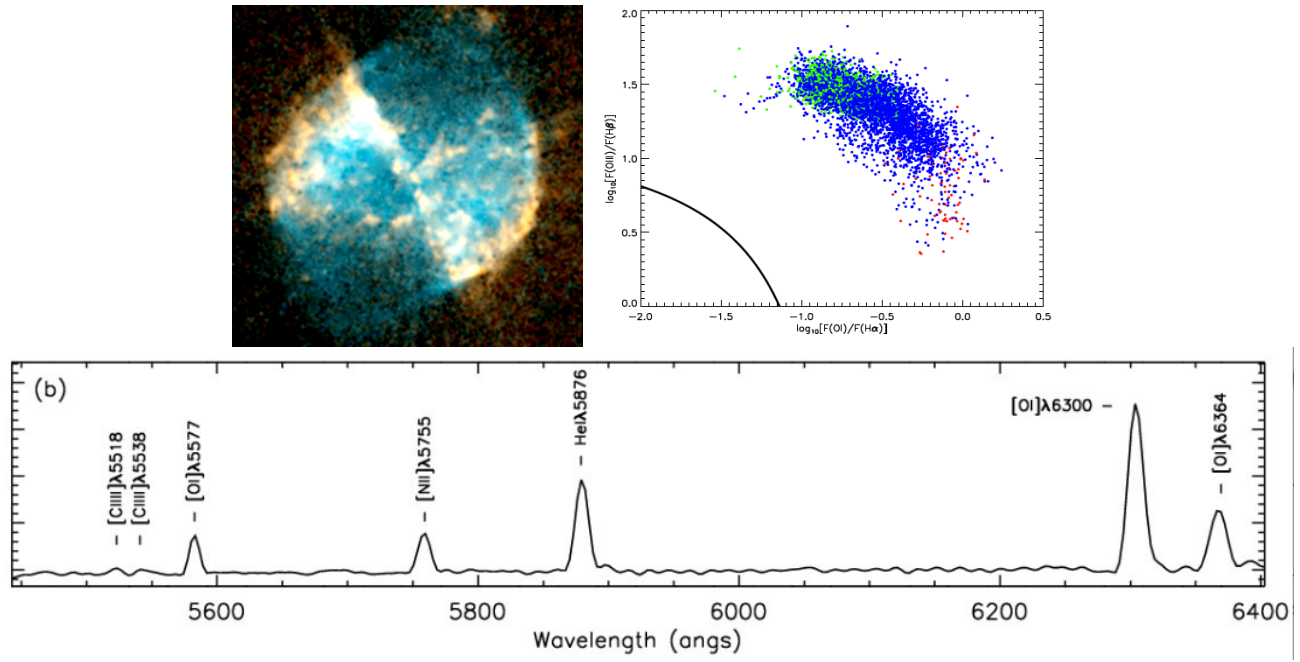

Fig. 2 (Upper Left) - Color composite image of the core of M27 in the lines of OI 6300 (orange) and HeI 5876 (blue). (Upper right) - BPT diagram of M27: log ([OIII] 5007 / Hβ) versus log ([OI]/Hα. (Lower panel) - Spectrum corresponding to a 5 x 300 arcsec "slit" extracted from the green data cube, after a careful sky subtraction. Note the presence of the auroral [NII] 5755 line, used to determine the electron temperature, as well as the weak [Cl III] doublet, whose ratio provides an estimate of the electron density of the nebula.

Early results, emphasizing the bright nebular lines, were presented by Drissen et al. (2008)[3], and we have recently obtained a data cube in the 545 - 635 nm region, where many weaker lines are present. This spectral region is also very rich in strong sky lines, resulting from light pollution (HgI 5461, 5770, 5791; NaI 5683, 5688, 5890, 5896) or the upper atmosphere ([OI] 5577, 6300, 6363), the latter being coincident with nebular lines. Because a significant number of pixels in the field of view were free of lines from M27 itself, a good determination of the average sky was possible, which was then subtracted from each pixel's spectrum (see details in Lagrois et al., in preparation [7]).

The auroral [NII] 5755 line can then be used in conjunction with the [NII] 6548, 6584 doublet as diagnostic for temperature measurements. The low excitation [OI] 6300, 6364 and 5577 lines provide a good estimate of electron temperatures prevailing in the $O^0$ volume. Finally, the [ClIII] 5518 / [ClIII] 5538 flux ratio is used to determine the electron densities in the high-excitation zone of the nebula. However, as can be seen in Figure 2, these lines are very weak and our estimate will be plagued with large uncertainties. SITELLE, when used at the CFHT, will greatly improve on SpIOMM's data in this regard.

## 2.3 Targeting the near-ultraviolet

The vast majority of Fourier transform spectrometers are used in the infrared. Indeed, a key ingredient to consider in the choice of a spectrometer, other than is global throughput, is its ability to decipher the spectral features from the incoming light. For an FTS, this property is quantified by the modulation efficiency (ME) of its central component, the interferometer. Since one of the most important contributors to the ME is the surface quality of the optical components (mirrors and beamsplitter), the modulation efficiency, for a given mirror quality, rapidly degrades at short wavelengths. SpIOMM's mirrors and beamsplitter were manufactured at λ/20 in the visible band. At 632 nm, where the mirror alignment and the servo loop quality are measured and optimized at the beginning of each night, ME reaches 85%. This value however decreases to ~ 30% at 350 nm. Combined with the lower quantum efficiency of the CCDs in the near UV, SpIOMM is not particularly efficient in this wavelength range. SITELLE has been specifically designed to reach a much higher ME in the blue and near-UV.

Nevertheless, we have obtained a few data cubes using the SDSS u' filter in order to observe the important diagnostic [OII] line, including one of the spiral galaxy M101. Another target was the bright planetary nebula M97, for which blue and red cubes were already in hand. The results, shown in Figure 3, are encouraging, since both the [OII] 3727 and the [NeIII] 3869 lines are easily detected and show different morphologies.

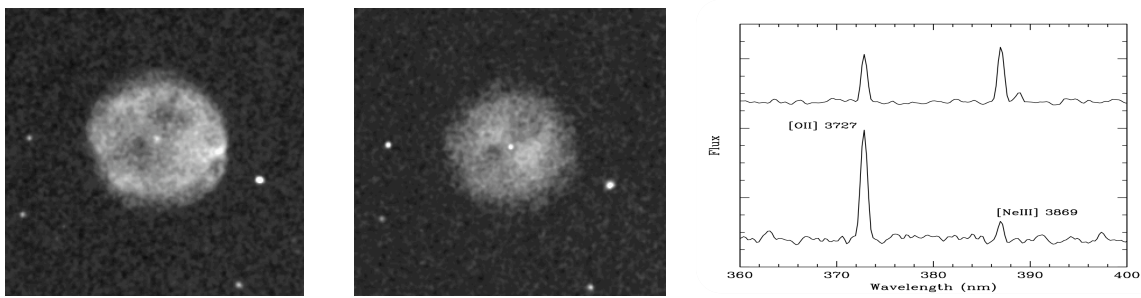

Fig. 3 - Images ([OII] 3727 on the left, [NeIII] 3969 on the right) and spectra of two different regions of the planetary nebula M97 extracted from an SDSS-u' data cube obtained with SpIOMM. Note the different morphologies (and hence line ratios) of the nebula in the two emission lines, indicative of its ionization structure.

## 2.4 Targeting absorption features

Contrary to a dispersive spectrograph, where photon noise at a given wavelength is due to the number of photons at this particular wavelength, FTS suffer from their "multiplex disadvantage": because the flux from the entire waveband (defined by the filters) hits the detector at a given position at each and every step of the interferogram, photon noise is distributed: it is caused, at a given wavelength, by all photons coming from all wavelengths. This is why imaging FTS are much more efficient than classical, dispersive Integral Field Spectrographs for the study of emission lines, but not so when the target's spectrum is composed of a continuum and absorption features. Nevertheless, the vast majority of FTS used for commercial applications and remote sensing of the Earth's atmosphere produce exquisite quality spectra of continuum and absorption sources.

Many mechanisms over the history of a galaxy may be considered to explain the compression of the gas responsible for bursts of star formation: external processes like galaxy interactions and extragalactic gas inflow, and internal processes like spiral arms and bar and feedback by stellar winds and supernovae. To understand the relative impact of these mechanisms, we have undertaken a study of the different stellar generations present in a sample of spiral galaxies. SpIOMM high spatial resolution and wide field of view, combined to stellar population synthesis codes, allow us to characterize the different stellar populations (in age, metallicity, and mass) and to study the hot gas that are sculpting the galaxies over time. While nebular emission lines are essential to probe the youngest generation of stars, absorption features from the stars themselves are a key ingredient to understand the more distant past of these galaxies.

Up until recently, SpIOMM was not used to study absorption lines. However, the installation of a camera to record the second output port, combined with improvements in the reduction software, have made absorption work possible. While SpIOMM records the spectra of dozens of stars while studying nebulae and galaxies, a better illustration comes from the spectrum of the core of the spiral galaxy NGC 3344 (Figure 4). The clear detection of the Ca + Fe absorption feature from the old stellar population in the nucleus of this galaxy, as well as the detection of other absorption lines (in particular Hβ) in other galaxies have paved the way to the study of stellar populations through the analysis of their absorption lines, first with SpIOMM but certainly with SITELLE at the CFHT.

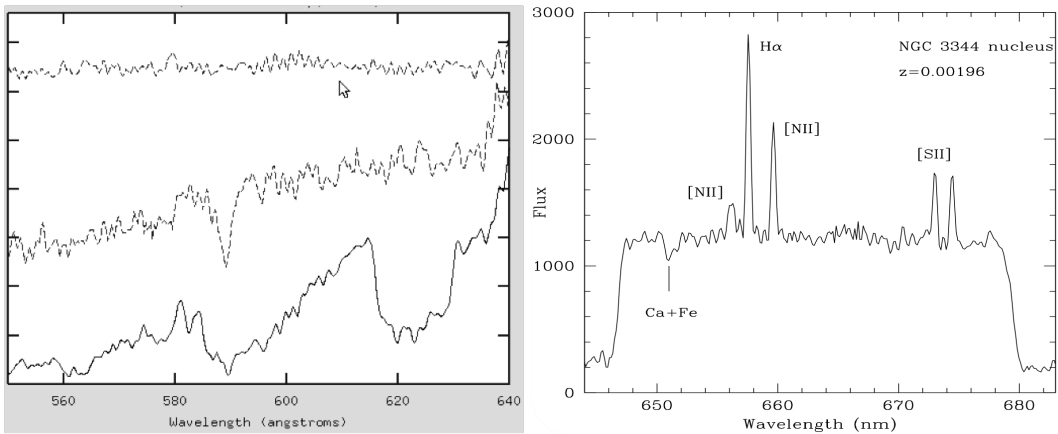

Fig. 4 - (Left) - Spectra of three stars (V ~ 14.8) in the field of the planetary nebula M27. From the top to the bottom: the central white dwarf (featureless), a G star (with the NaI 5890 absorption line) and an M star (with conspicuous TiO bands). (Right) - Spectrum of the nucleus of the spiral galaxy NGC 3344 (see image in Figure 6). Note the well-defined edges of the filter, the nebular lines (with particularly strong [NII] and [SII] lines), but more importantly the Ca + Fe absorption feature originating from the old stellar population.

## 2.5 Targeting chemical enrichment: the case of M1-67

Wolf-Rayet stars are the evolved descendants of the most massive stars. Analysis of their atmosphere clearly shows chemical enrichment with the products of the CNO nucleosynthesis (mostly He and N; WN stars) or helium burning (WC stars). Some of them are surrounded by a wind-blown bubble created as a cavity in the interstellar medium, which might also be chemically enriched. WR124 is a WN8h star surrounded by the nitrogen-rich M1-67 nebula, composed almost exclusively of stellar ejecta[8]. This nebula was observed with SpIOMM just a few weeks before this meeting, so the analysis is far from complete. However, nitrogen enrichment by at least a factor of five is obvious given the relative strength of the Hα, [NII] and [SII] lines. The morphology of the nebula in the Hα and [NII] lines is also quite different from one another, possibly indicating an inhomogeous ejection of chemically processed material by the stellar wind. It is a definite advantage of integral field spectrographs over long slits to cover large portions of an object with 100% filling factor.

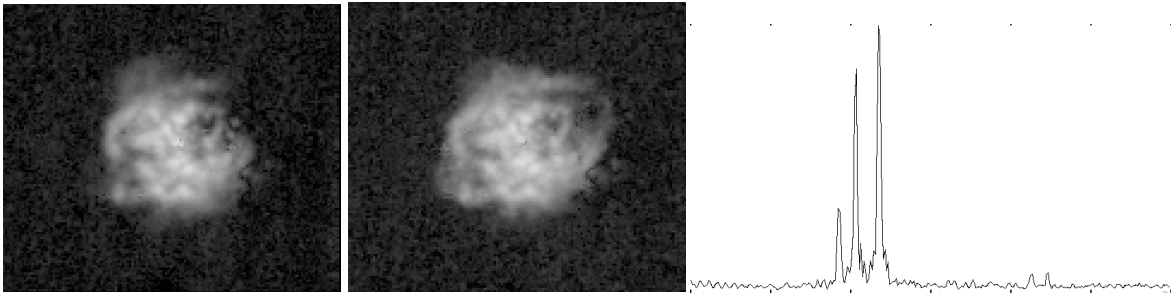

Fig. 5 - From left to right: Hα and [NII] images of M1-67, and a 4" x 4" spectrum in the 650 - 680 nm region near the star itself. Note the strength of the [NII] 6548, 6584 doublet on each side of Hα, as well as the weak [SII] 6717, 6731 doublet.

## 2.6 Targeting more distant galaxies

One of the clear advantages of SpIOMM over other IFS is its wide field of view. It is therefore better suited to study large, nearby galaxies than more distant ones which cover a small fraction of the detector's field of view. However, we have targeted a sample of interacting galaxies in order to study their kinematics and chemical abundances as well as to detect possible tidal tail dwarf galaxies resulting from the interaction and dispersed over a wide area.

Compact groups of galaxies have an special interest in extragalactic astrophysics since they are aggregates of (few) galaxies where interactions have revealed to play a very important role given the short projected distances and low velocity dispersions among their members.

We show in Figure 6 a comparison between the nearby, unperturbed, spiral NGC 3344 and the interacting pair Arp 84. Note the unusual distribution of HII regions in the latter.

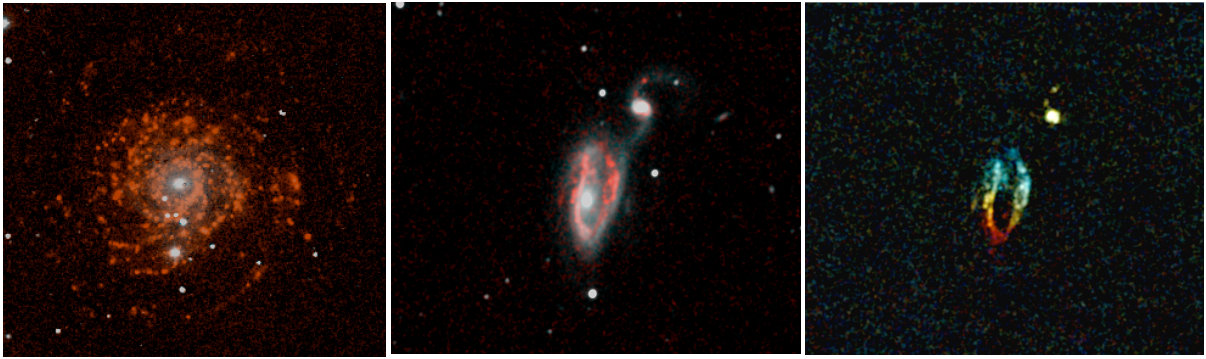

Fig. 6 - From left to right: Hα (in red) + red continuum (in white) images of the nearby spiral NGC 3044 (8' x 8') and the more distant interacting pair Arp 84 (3' x 3'), as well as the Doppler image of Arp 84 from the Hα and [NII] nebular lines.

## 3. CONCLUSION

In a thorough review of the imaging FTS properties, Maillard et al. [9] demonstrate, on paper, that this type of instrument has an important niche to fill among integral field spectrometers. By carrying on observations of a variety of astronomical targets with SpIOMM, we have demonstrated not only the feasibility of building and operating an astronomical imaging FTS working in the visible range (which remains, to our knowledge, unique in the world) but alos the interest of imaging FTS as wide-field spectrometers, at least in the niche of medium spectral resolution. The next steps with SpIOMM will involve the observation of star clusters at medium resolution, of bright nebulae with higher spectral resolution ($R \sim 10000$), as well as very low resolution ($R \sim 50 - 100$) of star clusters and more distant galaxies.

While we are presently designing and building SITELLE (Grandmont et al., these proceedings), an imaging FTS for the Canada-France-Hawaii telescope, all the experience gained with SpIOMM over the years has been absolutely essential, both to understand the limits and capabilities of such imaging spectrographs and to develop data reduction and analysis tools.

## ACKNOWLEDGEMENTS

We acknowldege funding by the Canadian Foundation for Innovation, Québec's FQRNT, Canada's NSERC, and Université Laval. We would like to thank ABB for its continuing support for the development of astronomical imaging FTS.